\documentclass[aps,prf,11pt,superscriptaddress,floatfix,tightenlines,showpacs,longbibliography,notitlepage]{revtex4-2}
\usepackage[normalem]{ulem}
\usepackage{soul}
\usepackage{xcolor}
\usepackage{amsfonts}
\usepackage{graphicx}% Include figure files
\usepackage{dcolumn}% Align table columns on decimal point
\usepackage{bm,amsmath,amssymb}% bold math
\usepackage{siunitx}
\usepackage[breaklinks,colorlinks = true,linkcolor = blue,urlcolor=blue,citecolor=blue]{hyperref}
\AtBeginDocument{%
    \newwrite\bibnotes
    \def\bibnotesext{Notes.bib}
    \immediate\openout\bibnotes=\jobname\bibnotesext
    \immediate\write\bibnotes{@CONTROL{REVTEX42Control}}
    \immediate\write\bibnotes{@CONTROL{%
    apsrev42Control,author="08",editor="1",pages="1",title="0",year="1"}}
     \if@filesw
     \immediate\write\@auxout{\string\citation{apsrev42Control}}%
    \fi
}%

\begin{document}

%\preprint{APS/123-QED}

\title{%When are airways plugged by annular mucus films?\\
% How wavelength-selection and sliding determine plugging of airways}% 
Probabilistic plugging of airways by sliding mucus films
}
% Force line breaks with \\
\author{Swarnaditya Hazra}
 % \altaffiliation[Also at ]{Department of Chemical Engineering, Indian Institute of Technology Bomaby}%Lines break automatically or can be forced with \\
\author{Jason R. Picardo}
 \email{picardo@iitb.ac.in}
\affiliation{Department of Chemical Engineering, Indian Institute of Technology Bombay, Mumbai 400076, India}
%  This line break forced with \textbackslash\textbackslash
% }%

% \collaboration{MUSO Collaboration}%\noaffiliation

%  \homepage{http://www.Second.institution.edu/~Charlie.Author}
% \affiliation{
%  Second institution and/or address\\
%  This line break forced% with \\
% }%
% \affiliation{
%  Third institution, the second for Charlie Author
% \affiliation{%
%  Authors' institution and/or address\\
%  This line break forced with \textbackslash\textbackslash
% }%

% \collaboration{CLEO Collaboration}%\noaffiliation


%Supplement Links:
% https://bighome.iitb.ac.in/index.php/s/nLdG89KST55DSpX

\begin{abstract}

When do mucus films plug lung airways? Using reduced-order simulations of a large ensemble of randomly perturbed films, we show that the answer is not determined by just the film's volume. While very thin films always stay open and very thick films always plug, we find a range of intermediate films for which plugging is uncertain. 
% being the outcome of dynamic processes that are sensitive to the random perturbations that trigger the Rayleigh-Plateau instability. 
The fastest-growing linear mode of the Rayleigh-Plateau instability ensures that the film's volume is divided among multiple humps. However, the nonlinear growth of these humps can occur unevenly, due to spontaneous axial sliding---a lucky hump can sweep up a disproportionate share of the film's volume and so form a plug. This sliding-induced plugging is robust and prevails with or without gravitational and ciliary transport.
\end{abstract}

\maketitle

%\tableofcontents

\textit{Introduction.} Lung airways are lined by a thin film of mucus, whose free surface is susceptible to the Rayleigh-Plateau instability \citep{Levy2014}. In an idealized tubular airway, of radius $R$, an axisymmetric and initially-uniform mucus film spontaneously gathers itself into two distinct structures: unduloids and liquid-bridges \citep{everett1972model}. The former are equilibrium-shaped annular humps which leave the airway open for the passage of airflow. The latter, also called mucus plugs, occlude the airway \citep{Romano2019viscous} and can lead to respiratory distress, as occurs in rapid-onset fatal asthma \citep{hays2003role,rogers2004airway}. It is therefore important to know when the mucus film forms a plug. 

\begin{figure}
\centering
\includegraphics[width=1.0\textwidth]{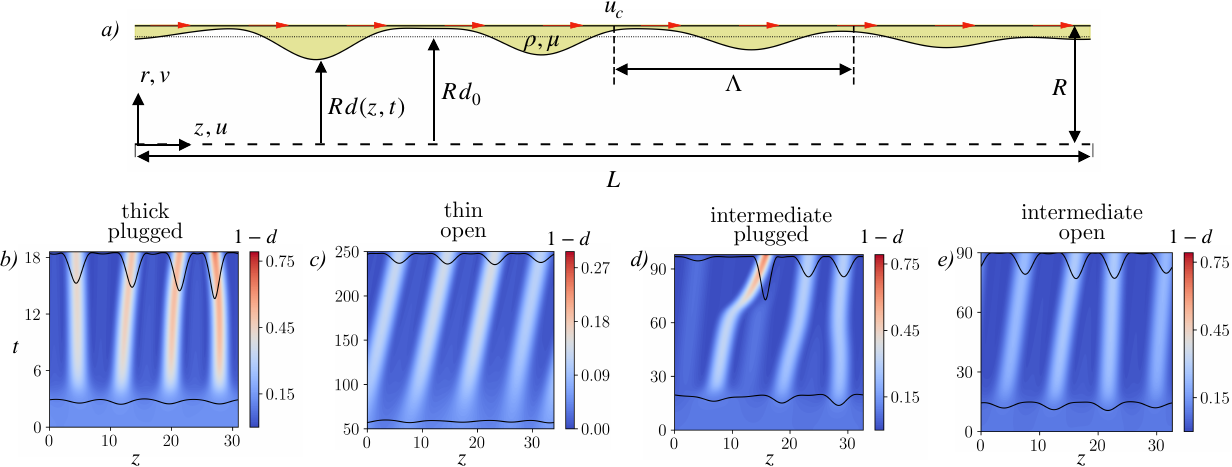}% Here is how to import EPS art
\caption{\label{fig:kymo} 
(a) Schematic of the annular mucus film in a long axisymmetric airway, spanning multiple wavelengths of the fastest-growing mode $\Lambda$. (b-e) Kymographs of the evolution of the film, in representative randomly-initialized runs, for thick, thin, and intermediate films: 
% (b) $1-d_0 = 0.13$, all runs lead to plugging; (c) $1-d_0 = 0.047$, all runs remain open; (\textit{d-e}) $1-d_0 = 0.081$, plugging occurs in $?\%$ of the runs. 
(b) $1-d_0 = 0.13$, (c) $1-d_0 = 0.047$, (\textit{d-e}) $1-d_0 = 0.081$.
% Panels (b), (c), (d), and (e) are animated in supplementary
Panels (b-e) are animated in supplementary \href{https://bighome.iitb.ac.in/index.php/s/nf9eqr2G5cbp9ZF}{movie 1}, \href{https://bighome.iitb.ac.in/index.php/s/w9X3o8BE4zwGdjE}{movie 2}, \href{https://bighome.iitb.ac.in/index.php/s/pgDxQAEHLz6wHpM}{movie 3} and \href{https://bighome.iitb.ac.in/index.php/s/S34zeEr7Bab76Xf}{movie 4} \citep{supplement} 
% Panel \textit{?} is animated in two parts in \href{https://bighome.iitb.ac.in/index.php/s/gQGtdSmSbQWrPtX}{movie 4} and \href{https://bighome.iitb.ac.in/index.php/s/7k7QPYtcsS9msA9}{movie 5}.
}
\end{figure}

Both unduloids and plugs are equilibrium shapes of constant mean curvature.
% , i.e. their axial and radial curvatures maintain the same balance at every point along their profile. 
However, unduloid solutions exist only for liquid volumes below $1.73\pi R^3$ \citep{everett1972model,Dietze2015}, after which they give way to plugs. Naively applying this volume criterion to an airway of length $L$, with an initially uniform film of thickness $(1-d_0)R$ (see figure~\ref{fig:kymo}a, in which $d_0R$ is the distance of the flat film from the centreline), leads to the conclusion that the airway would get plugged if $(1-d_0^{2}) > 1.73R/L$.
% , i.e. if the initial nondimensional thickness $1-d_0$ exceeds $1-(1-1.73 R/L)^{1/2}$.  
But this implies that, to avoid plugging, the film must become ever thinner ($d_0$ approaching 1) as the airway length $L$ is increased. Of course, in reality, an unstable film in a long airway does not gather itself into just one large hump, but rather organises itself into several growing humps. The question then is how many humps emerge from the initially uniform film, as a result of the Rayleigh-Plateau instability, and whether all of them are able to saturate into stable unduloids. Thus, as shown in this letter, the formation of plugs
% by thin films on long domains 
is governed not solely by equilibrium considerations but also by interfacial pattern selection and nonlinear film dynamics.
% dynamics of the draining film. 

% At low volume fractions, the film accumulates into unduloids, i.e. equilibrium-shaped annular humps or collars, which are separated by depleted zones that are devoid of liquid. At higher volume fractions, unduloid solutions cease to exist and instead the film forms liquid bridges \citep{everett1972model}. Not only do these bridges block the airway, their formation is associated with strong recoil forces which can damage the cells of the airway wall~\citep{Romano2019viscous,erken2022capillary}. Several studies have therefore analysed liquid-bridge formation, considering the effects of mucus rheology~\citep{Romano2021viscoelastic,Romano2023viscoplastic} and surfactants~\citep{romano2022effect}. If the wall is soft, as in the terminal airways, closure can occur even at lower volume fractions, because the capillary-induced reduction of pressure inside liquid humps causes the wall to collapse~\citep{heil2008mechanics}. Here, we are interested in non-collapsing, rigid-walled airways with mucus fractions that are low enough to avoid closure.

Asking how many humps will emerge from the unstable annular film is akin to asking how many drops will arise from an unstable liquid jet. The latter was addressed by Rayleigh, in his seminal work on pattern selection \citep{Rayleigh1879}, in which he proposed that the fastest-growing instability mode will dominate the dynamics and enforce its wavelength $\Lambda$ onto the pattern of emerging interfacial undulations \citep{narayanan}. This principle of wavelength selection works quite well in subcritical interfacial instabilities, like Rayleigh-Plateau and Rayleigh-Taylor (with the exception of special circumstances in which the linear-growth-rate curve exhibits multiple peaks \citep{Picardo2017pattern}). The perfect realization of this paradigm in the present problem would have the initially-uniform film of length $L$ organize itself into identical repeating units of length $\Lambda$. The liquid volume contained in each unit cell would then give rise to a growing hump, which would saturate and form a stable unduloid if 
% $ (1-d_0) < 1.73R/\Lambda (1+d_0)$. 
$ (1-d_0^2) < 1.73R/\Lambda$. 
Noting that $\Lambda \approx 2^{3/2}\pi  d_0 R$ (this inviscid result of \citet{Rayleigh1892wavelength} works very well even for viscous films \citep{Dietze2015}), we see that plugs would form only if the initial thickness exceeds $1-d^{\Lambda}_0 \approx 0.117$ [obtained by solving $1-(d^{\Lambda}_0)^2 = 1.73 R/ \Lambda = 2^{-3/2}(1.73) /\pi d^{\Lambda}_0 $ and discarding the small $d^{\Lambda}_0$ solution that is irrelevant to thin films]. So, accounting for the emergence of multiple humps suggests that much thicker films will remain open in long airways, because the critical thickness no longer scales as $L^{-1}$ but becomes independent of $L$. 

% Thus, accounting for pattern selection yields a critical thickness that is independent of the length $L$ of the long airway. So, for situations where $L \gg \Lambda$, the dominance of the fastest growing mode and the consequent emergence of multiple humps should allow for open airways with much thicker films because the critical thickness will not scale as $L^{-1}$. 

To test these ideas, we perform numerous simulations on long domains, with $L = 4\Lambda$, and examine the evolution of randomly perturbed films with a range of initial thicknesses, $1-d_0$.
% we randomly perturb the uniform film and follow its evolution until either plugging occurs or an array of stable unduloids are formed. By simulating an ensemble of independent realizations, for each value of $1-d_0$, we obtain statistics on plugging as a function of film thickness. 
On the one hand, we show that films thicker than the $L$ based critical thickness of $1-d^L_0\approx 0.025$ [which satisfies by $1-(d^L_0)^2 = 1.73 R/L =  1.73 R/(4 \Lambda)$] do remain open due to the emergence of multiple humps. On the other hand, we find that films thinner than $1-d^{\Lambda}_0 = 0.117$ can plug the airway due to \textit{nonlinear} film dynamics that violate the criterion based on the fastest-growing \textit{linear} mode. Specifically, the sliding instability of growing humps \citep{Dietze2018sliding} can allow a lucky hump to sweep up more than its 1/4th share of the film's volume and thus exceed the unduloid limit.
% , even though the entire film satisfies $ (1-d_0) < 1.73/(1+d_0)\Lambda$.

\textit{Model.} Our model airway consists of a cylindrical tube lined with an annular mucus film (figure~\ref{fig:kymo}a). For simplicity, we assume that the mucus is Newtonian and that the air phase is passive. This setting is realized to a good approximation in mid-generation conducting airways, where the air flow is laminar and relatively weak \citep{pedley1977pulmonary} and does not affect the  evolution of the mucus film \citep{Swarnaditya-particles}. The airway walls are relatively rigid, unlike the terminal airways which collapse under capillary pressure differences \citep{heil2008mechanics}. The wall are also ciliated, and so we include mucociliary transport via a simple coarse-grained boundary condition (we show later, though, that our conclusions are unaffected by ciliary transport). The wall-attached cilia reside beneath the mucus film, within a layer of watery periciliary liquid (PCL)~\citep{Sleigh1988,Boucher2002,Boucher2006}. The synchronous and asymmetric beating of cilia
% ---only the upright forward stroke engages the mucus film directly---
transports the mucus film at a rate of about $25-60$ \si{\mu m.s^{-1}} \citep{Levy2014,Dietze2023mucociliary}. 
While several cilia-resolving simulations of mucociliary transport have been developed \citep{Smith2007Stokeslet,smith2018Blaketribute,Sedaghat2016lbm}, 
the study of large-scale mucus flows necessitates a coarse-grained approach in which ciliary forcing is modelled via a boundary condition at the base of the mucus film \citep{vasquez2016modeling,Bottier2017exp,Bottier2017model}. 
The simplest version prescribes the mucus velocity at the wall in terms of a travelling wave function, $u_c$, which mimics the metachronal beating of cilia. However, for a Newtonian film, the rapid oscillations of the travelling wave do not alter the slow transport of the film and so it can be replaced by just its constant spatial (or temporal) average \citep{Swarnaditya-particles,Dietze2023mucociliary}. While one could also include Navier slip to account for the lubricating effect of the PCL layer \citep{Bottier2017model,Dietze2023mucociliary}, we here opt for the simple Couette condition which is sufficient to capture the primary effect of cilia, namely the axial transport of the radially-deforming film.

To facilitate simulations on long domains, we use a reduced-order thin film model based on the weighted residual integral boundary layer (WRIBL) approach of \citet{Dietze2015}. This model consistently incorporates the effects of nonlinear interface curvature---crucial for capturing airway plugging---and second-order longitudinal viscous stresses, which are important near thinning necks. The WRIBL model has been extensively validated against direct numerical simulations of the Navier-Stokes equations \citep{Dietze2013,Dietze2015}. In the case of a passive core-fluid, the flow of the mucus film (of viscosity $\mu$ and density $\rho$) is described by two coupled partial differential equations for the evolution of the interface (endowed with interfacial tension $\gamma$), located at $r = d(z,t)$, and the radially-averaged flow rate $2 \pi Q(z,t)$:
% \vspace{-\baselineskip}
\begin{equation}\label{kbc_mucus}
   d\, \partial_t d = \partial_z Q,  
\end{equation}
\vspace{-2\baselineskip}
% \begin{multline}\label{wribl_q_eq_pas}
% Re \bigl( {S}_{m}{\partial_t Q}+ {S}_{c}{\partial_t u_c}+ {F}_{m} Q\partial_z Q +  {F}_{c} u_c\partial_z Q+ {H}_{m}Q\partial_z u_c+ {H}_{c}u_c\partial_z u_c+\\
% + {G}_{mm}Q Q\partial_z{d}+ {G}_{mc}Q u_c\partial_z{d}+ {G}_{cc}u_c u_c\partial_z{d}\bigr)\\
%  =Ca\partial_z\kappa  {Iw}+Q+ {J}_{m}Q(\partial_z {d})^2+ {J}_{c}u_{c}(\partial_z {d})^2+ {K}_{m}\partial_z Q\partial_z {d}\\+ {K}_{c}{\partial_z u_{c}}{\partial_z {d}}
%  + {L}_{m}Q{\partial_z^2 {d}}+ {L}_{c}u_{c}{\partial_z^2 {d}}+ {M}_m{\partial_z^2Q}+ {M}_c{\partial_z^2u_c}-u_c\partial_r{ {w_m}}|_{1}
% \end{multline}
% \begin{multline}\label{wribl_q_eq_pas}
% Re \bigl( {S}{\partial_t Q}+ {F}_{m} Q\partial_z Q +  {F}_{c} u_c\partial_z Q+
% + {G}_{mm}Q Q\partial_z{d}+ {G}_{mc}Q u_c\partial_z{d}+ {G}_{cc}u_c u_c\partial_z{d}\bigr)\\
%  =Ca\partial_z\kappa  {Iw}+Q+ {J}_{m}Q(\partial_z {d})^2+ {J}_{c}u_{c}(\partial_z {d})^2+ {K}_{m}\partial_z Q\partial_z {d}\\
%  + {L}_{m}Q{\partial_z^2 {d}}+ {L}_{c}u_{c}{\partial_z^2 {d}}+ {M}_m{\partial_z^2Q}-u_c\partial_r{ {w_m}}|_{1}
% \end{multline}
\begin{multline}\label{wribl_q_eq_pas}
Re \bigl[ {S_m}{\partial_t Q}+ \left({F}_{m} Q+ {F}_{c} u_c \right)\partial_z Q
+ \left({H}_{mm}Q^2+ {H}_{mc}Q u_c+ {H}_{cc}u_c^2\right) \partial_z{d}\bigr]
 =Q - {N_c u_c} +\mathcal{G} \\+ Ca\partial_z\kappa  {I}+ \left({J}_{m}Q+ {J}_{c}u_{c}\right)(\partial_z {d})^2+ {K}_{m}\partial_z Q\partial_z {d}
 + \left({L}_{m}Q+ {L}_{c}u_{c} \right){\partial_z^2 {d}}+ {M}_m{\partial_z^2Q},
 % {u_c\partial_r{ {w_m}}|_{1}},
\end{multline}
where $\kappa =d^{-1}-\partial_{zz}d-(\partial_z d)^{2}(2d)^{-2}$ is a second-order long-wave approximation to the interface curvature. The subscript-bearing coefficients, $S_m$, $F_m$, $F_c$, and so on, are functions of $d$. 
% Equations \eqref{kbc_mucus} and \eqref{wribl_q_eq_pas} 
These equations have been rendered dimensionless using $R$, $\tau_{RP}$, and $R/\tau_{RP}$, where the time scale $\tau_{RP} =  ({\mu R}/{\gamma}) 24 (d^{\Lambda}_0)^5[(1-d^{\Lambda}_0)^2(1-(d^{\Lambda}_0)^2)]^{-1}$ is the inertialess approximation to the inverse growth rate of the fastest-growing mode \citep{johnson1991,Dietze2015}, for a film of thickness $1-d^{\Lambda}_0 = 0.117$.
% , which satisfies $(1-(d^{\Lambda}_0)^2)=1.73/2\pi2^{1/2}d^{\Lambda}_0$ (where we have approximated $\Lambda$ as $2\pi2^{1/2}d_0 R$).
We use parameter values that are relevant to mucus films in the middle airways: $\mu = 0.01$ \si{Pa.s}, $\rho=1000$ \si{kg.m^{-3}}, $\gamma = 0.05$ \si{N.m^{-1}}, $u_c R/\tau_{RP} = 40$ \si{\mu m.s^{-1}}  \citep{Swarnaditya-particles} (except when we exclude cilia by setting $u_c = 0$). {Thus, $Re = {\rho R^2}/{\mu \tau_{RP} } = 0.047$ and $Ca ={\gamma \tau_{RP} }/{\mu R } = 4206.6$}. The dimensionless body force $\mathcal{G}$ is set to zero except when we include weak gravitation forcing. With these parameter values, $\Lambda \sim 3$ \si{mm}, while an airway's length is about 5 - 10\si{mm} \citep{Sleigh1988}; so it is indeed physiologically relevant to investigate the fate of a mucus film that is long enough to give rise to multiple humps. 
% $u_c$ is the ciliary velocity (set to zero when cilia is excluded) that is applied to the base of the film. 
% can act as an alternate source of axial asymmetry to ciliary forcing, so that our conclusions regarding sliding-induced plugging remain valid even without cilia. 

Equations \eqref{kbc_mucus} and \eqref{wribl_q_eq_pas} are similar to the passive-core model of \citet{Dietze2015}, except for the inclusion of the ciliary transport velocity $u_c$. We present a detailed derivation in \citet{Swarnaditya-particles}, where we also show that the evolution of the mucus film (i) is almost entirely unaffected by the oscillatory air flow and (ii) is left unchanged on replacing the travelling-wave form of $u_c$ by its constant mean value (as is done here). In the {supplemental material} \citep{supplement}, we present a validation of our model with the active-core model of \citet{Dietze2015}, along with their DNS results, for an annular mucus film.

% investigate the effect of weak gravitation forcing. 
 
% show that the axial asymmetry due to weak gravitational forcing is sufficient to trigger sliding so that our conclusions regarding plugging and its stochastic nature remain relevant even in the absence of ciliary transport. 

\textit{Simulations.} We numerically solve Eqs.~\eqref{kbc_mucus} and \eqref{wribl_q_eq_pas}, by discretizing the $z$ direction using a second-order finite-difference method. Though the domain length $L$ is meant to represent the entire airway's length, we still apply periodic conditions at the boundaries because they are the simplest conditions that are compatible with ciliary transport. Setting aside more realistic intlet/outlet conditions for future work, allows us to focus here on the intrinsic dynamics of the film. Time integration is performed using the stiff, adaptive time-stepping algorithm LSODA\citep{hindmarsh1995algorithms}, provided in the solve{\textunderscore}ivp function of the Python library SciPy \citep{scipy}. Such a procedure is easier to implement when evolution equations are available for both $d$ and $Q$, and so we retain the inertial terms in Eq~\ref{wribl_q_eq_pas}, despite the small value of $Re$. The supplemental material \citep{supplement} links to a Python Jupyter notebook that simulates the equations, after importing the lengthy analytical expressions of the coefficients from text files. 

% The initially-uniform film 
% of thickness $1-d_0$ 
% is subjected to a random perturbation:
% , of the form 
Using the initial condition, $ d(z,0) = d_0 + 10^{-3} \sum_{m=1}^{10} A_m\,\mathrm{cos}(2 \pi m z/L + \varphi_m)$,
% \begin{equation}
%     d(z,0) = d_0 + 10^{-3} \sum_{m=1}^{10} A_m\,\mathrm{cos}(2 \pi m z/L + \varphi_m),
% \end{equation}
we randomly perturb the initially uniform film of thickness $1-d_0$. Here, 
 $A_m$ and $\varphi_m$ are uniformly distributed random numbers between $[0,1]$ and $[0, 2 \pi]$, respectively, and $L = 4 \Lambda$ (unless specified otherwise). The film is evolved until either a quasi-static array of unduloid-like humps are formed, or one of the humps experiences runaway growth and approaches the centreline. Since our thin film model cannot resolve the coalescence event, we stop the simulation when $d \approx 0.1$ (our results are not sensitive to a further decrease of this threshold). We perform an ensemble of 80 realizations for each initial film thickness---totalling over a thousand simulations---and thus obtain statistics on the formation of plugs.

% \textcolor{red}{Mention in discussion of kymographs: that formed stable humps were run for a long time, of around $900T_b$, to ensure that the asymptotic state was indeed an open airway; for comparison, closure events typically occurred within $100T_b$ and often much faster (depending on the film's volume, \textit{i.e.}, on $d_0$).}

\textit{Illustration of plugged and open realizations.} Our results show that films fall into three categories based on their initial thickness: thick films that always plug the airway, thin films that never form a plug, and intermediate films which plug the airway with a probability that increases with their thickness. Representative examples of the evolution of the film for each of these cases is shown in the kymographs of Figs.~\ref{fig:kymo}(b)-(e). Figure \ref{fig:kymo}(b) corresponds to a thick film with $1-d_0 = 0.13$, which is well above the fastest-mode threshold of $1-d^\Lambda_0=0.117$. In this case, even after the film's volume is divided into four humps by the dominance of the fastest-growing linear mode, the volume within each hump exceeds the unduloid limit. It is not surprising then that all realizations of this thick film form a plug. In Fig. \ref{fig:kymo}(b), the right-most hump is favored by the initial perturbation and approaches the centerline first (see the animation in \href{https://bighome.iitb.ac.in/index.php/s/EFB2atpLw4NrQcG}{movie 1} of the supplemental material \citep{supplement}). 
% A plug is therefore guaranteed to form when $1-d_0 > 0.117$. 

Well below the threshold of $1-d^\Lambda_0=0.117$, but still above the $L$ based critical thickness of $1-d^L_0=0.025$, the film is sufficiently thin for all humps to form stable unduloids and so plugging never occurs. An example is seen in Fig.~\ref{fig:kymo}(c) which corresponds to $1-d_0 = 0.047$ (visualized in \href{https://bighome.iitb.ac.in/index.php/s/EFB2atpLw4NrQcG}{movie 2} \citep{supplement}); here the slow and uniform translation of humps is due to the ciliary velocity {$u_c = 0.034$}. 
% [note the difference in the time axes of Figs. \ref{fig:kymo}(b) and \ref{fig:kymo}(c)]. 
In such open cases, the mucus in the thin depleted zones continues to drain into the adjacent humps, so that  
% such that the film thickness in these regions \textcolor{red}{decreases algebraically \citep{lister2006capillary}}. So eventually, in finite time, 
the interface will eventually make contact with the wall, in finite time \citep{lister2006capillary}. The depleted zones will subsequently dry out leaving behind isolated unduloids. As our thin-film model cannot simulate dryout (without the added complication of a precursor film), 
% and because our focus here is on plugging, 
we stop the simulations once the film thickness drops below 0.001. 
% We have checked that the specific value of this small threshold does not affect our conclusions. 
We have ensured that our spatial resolution is sufficient to resolve the flow in these thin depleted zones---the supplemental material shows that doubling the resolution leaves the plugging statistics unchanged \citep{supplement}. 

Films with initial thickness near $1-d^\Lambda_0=0.117$ exhibit more complex dynamics: some random realizations plug the airway while others remain open. These two outcomes are illustrated in Figs. \ref{fig:kymo}(d) and \ref{fig:kymo}(e), both for $1-d_0 = 0.098$ (see also \href{https://bighome.iitb.ac.in/index.php/s/pgDxQAEHLz6wHpM}{movie 3} and \href{https://bighome.iitb.ac.in/index.php/s/S34zeEr7Bab76Xf}{movie 4}  \citep{supplement}). In Fig. \ref{fig:kymo}(d), the rapid growth of the hump that ultimately plugs the airway is accompanied by a sudden and prominent sliding motion. A consequence of a secondary instability \citep{Dietze2018sliding}, such sliding was first reported by \citet{lister2006capillary} and has since been observed in several studies of thin films draining due to the Rayleigh-Plateau and Rayleigh-Taylor instabilities \citep{lister2006capillary,lister2006drop,glasner2007,Dietze2015,pillai_2018_electro,pillai_2018_evap}.

Sliding does occur even in the open realization, as evidenced by the slight change in the inter-hump spacing in Fig.~\ref{fig:kymo}(e). [Unlike sliding, the axial motion due to the cilia velocity $u_c$ maintains the inter-hump spacing, as seen in Fig.~\ref{fig:kymo}(c).] However, no single hump in the open realization slides much more than the others [Fig.~\ref{fig:kymo}(e)].  The sliding of unduloid-shaped humps in open tubes was analyzed in detail by \citet{lister2006capillary}, who showed that such humps, termed collars, can collide; plugging does not follow though because they always rebound without merging \citep{lister2006capillary}.

% The prominence of such dramatic sliding events decreases as the film thickness reduces and so the probability of closure also decreases (see figure~\ref{fig:Pf}a). 

\begin{figure}
\centering
\includegraphics[width=0.8\textwidth]{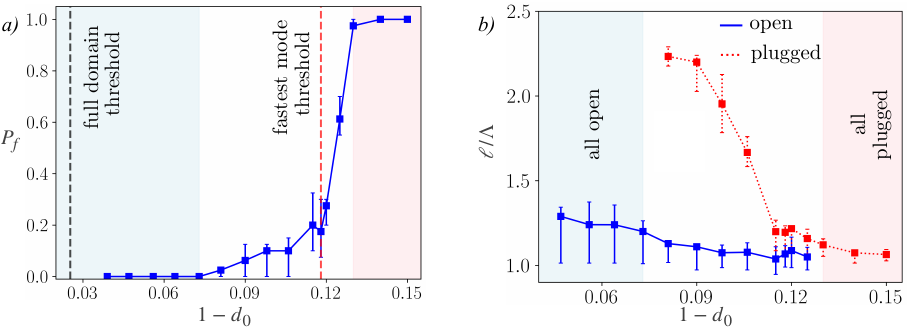}% Here is how to import EPS art
 \caption{\label{fig:Pf} (a) Probability of plugging 
 % (fraction of realizations that form a plug) 
 as a function of the initial film thickness. 
% The markers and bars represent the mean and inter-quartile range obtained from the ensemble of simulations for each $d_0$. 
The dashed vertical lines show the critical thickness beyond which unduloids do not exist, considering the film's volume in the full domain ($1-d^L_0 = 0.025$) and in the $\Lambda$ unit-cell ($1-d^\Lambda_0 = 0.117$). (b) The distance of the deepest hump from the farther of its two neighbors $\ell$, scaled by $\Lambda$ and plotted separately for open and closed realizations.}
\end{figure}

\textit{Probabilistic plugging.} Having seen some representative examples of the film's evolution, let us now turn to the statistical picture that emerges from the ensemble of simulations. Figure \ref{fig:Pf}(a) presents the fraction $P_f$ of realizations that end in a plugged airway, as a function of the initial thickness $1-d_0$. The bars here represent the uncertainty in the value of $P_f$. (For each value of $1-d_0$, we divide the ensemble of 80 simulations into groups of 10 and thus obtain 8 measurements of $P_f$; the corresponding mean and inter-quartile-range are presented as markers and bars; the mean is of course equal to the plugging fraction of the entire ensemble of 80 simulations.) We see, in Fig. \ref{fig:Pf}(a), a clear division into thick films that nearly always plug ($P_f \approx 1$; red region), thin films that never plug ($P_f = 0$; blue region), and intermediate films that plug with a probability $P_f$ that increases with their thickness (white region).
% as the threshold $1-d^\Lambda_0=0.117$ is approached. 

The finding that films thicker than $1-d^L_0 = 0.025$ always remain open [the blue region extends well beyond the full domain threshold in Fig. \ref{fig:Pf}(a)] shows that the randomly perturbed film always gives rise to multiple humps. That the probability of plugging increases rapidly once the thickness exceeds the fastest mode threshold $1-d^\Lambda_0=0.117$ shows that the fastest-growing linear mode governs the pattern of emerging humps. The small gap between the fastest-mode threshold and the regime of guaranteed plugging, in Fig. \ref{fig:Pf}(a), is likely a consequence of the inability of the thin film model to simulate the dynamics past dryout. The films in this gap always give rise to $L/\Lambda=4$ humps and at least one of them is sure to exceed the unduloid volume-limit, provided the film drains \textit{entirely} into the humps. However, our simulations stop the moment the interface meets the wall, at any point along the domain. So, plugs that might have formed by further drainage of the film are now missed. Therefore, we expect future work that accurately treats dried-out spots---and the moving contact lines that accompany their growth---to find that plugs are sure to form just beyond the fastest-mode threshold.

The most intriguing aspect of Fig. \ref{fig:Pf}(a) is that intermediate films, thinner than the fastest-mode threshold, do form plugs. This occurs because the four emerging humps do not evolve identically, and when one of them takes up much more than its $1/4^\mathrm{th}$ share of the film's volume, it can exceed the unduloid limit and plug the airway. The question, though, is how the lucky hump manages to accumulate so much excess volume.

\textit{Sliding-induced plugging.} On examining the evolution of several films in the intermediate region, we find that the key difference between the closed and open realizations is that, in the latter, no single hump undergoes excessive sliding [as illustrated in Figs. \ref{fig:kymo}(d) and \ref{fig:kymo}(e)]. As a simple way to quantify this observation, we calculate the distance $\ell$ of the deepest hump from the farther of its two neighbors [\textit{i.e.}, we measure the distance between the corresponding minima of $d(z,t)$]. We perform this measurement at the end of the simulations---just before plug formation, in case of plugged realizations, and just before the depleted film dries out at the wall, in case of open realizations. Now, if sliding either does not occur at all or occurs uniformly for all humps, then the spacing between humps will not change during the film's evolution and so should be near the wavelength of the fastest-mode $\Lambda$. In contrast, if one humps slides much more than the others, then $\ell$ should be greater than $\Lambda$. So, if preferential sliding is responsible for plugging of intermediate films, $\ell/\Lambda$ should be greater than unity for plugged realizations and close to unity for open realizations. This is indeed what we find in Fig. \ref{fig:Pf}(b), which shows that $\ell/\Lambda$ is significantly larger in  plugged cases. The variation with film thickness is explained below. 
% These plugged values approach those in open cases as the film thickness nears the fastest-mode threshold; this is to expected because, for thicker films, the lucky hump need not slide as much since it needs less extra-volume to form a plug. 

\begin{figure}
\centering
\includegraphics[width=0.8\textwidth]{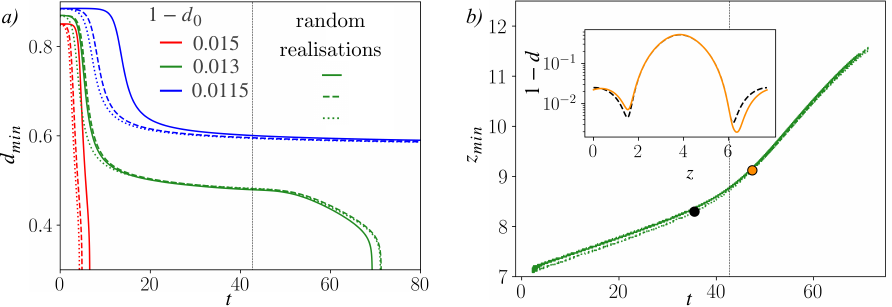}% Here is how to import EPS art
\caption{\label{fig:single} (a) Evolution of the minimum interface position $d_{min}$ for randomly perturbed films (three realizations) of different initial thicknesses on a short domain with a single growing hump ($L = \Lambda$). (b) Time-trace of the axial position of $d_{min}$ for the realizations of the intermediate-thickness film; the semi-log inset shows the film's profile in one of these realizations (animated in \href{https://bighome.iitb.ac.in/index.php/s/8Mj9NLDDPpN5XWo}{movie 5} \citep{supplement}) just before and after sliding.}
\end{figure}

To appreciate how sliding enables a hump to rapidly increase its volume, it is useful to consider the evolution of the film on a small domain of $L = \Lambda$. Note, first, that because only one hump can emerge in this domain, the question of when a plug forms is settled by the limiting volume of unduloids. So films thicker (thinner) than $1-d^L_0=1-d^\Lambda_0=0.117$ will form a plug (remain open) regardless of the initial perturbation. This is illustrated in Fig.~\ref{fig:single}(a), where the evolution of the minimum interface position $d_{min}$ is plotted for three random realizations, for three film thickness---one below the threshold (blue line) and two above (green and red lines). Interestingly, while the two thicker films do indeed form plugs, as evidenced by the rapid decreased of $d_{min}$ towards zero, the thinner of the two (green line) undergoes a long quasi-static phase in which the hump grows very slowly. The sudden acceleration of its growth is caused by the sliding instability, as explained in \citep{Dietze2015,Dietze2018sliding} and illustrated in Fig.~\ref{fig:single}(b). Here, the main panel traces the axial location of $d_{min}$: the hump that translates with the ciliary velocity during its initial growth and quasi-static phase begins to move faster just as it starts to grow rapidly (the vertical dashed line is provided as a rough marker of the onset of sliding). The inset of Fig.~\ref{fig:single}(b) shows the film's profile (on a log-scale) just before and after sliding. During the quasi-static phase, the two symmetric thinning necks on either side of the hump restrict flow into the hump and slow its growth. With the onset of the sliding instability, the necks break symmetry and one of them opens up, allowing much faster entry of fluid into the hump. (The self-reinforcing nature of asymmetric perturbations at the necks is elucidated in \cite{Dietze2018sliding}.)

On a short, single-hump domain, sliding only hastens plug formation which would have otherwise taken a much longer time. On a long domain, though, sliding causes films to plug that would otherwise have remained open; hence the plugging of films thinner than the fastest-mode threshold in Fig.~\ref{fig:Pf}.  Sliding-induced plugging requires one lucky hump to slide faster than the others, though, because if all humps slide equally then they will take up equal volumes and an intermediate-thickness film would not plug. Now, the chance for one hump to slide dramatically should increase with the number of humps, and so the plugging fraction should increase with the airway length. The supplemental material shows that this is the case by comparing results for $L =3 \Lambda$ and $4\Lambda$ \citep{supplement}.
Also, for thicker films, the lucky hump would have to accumulate less excess volume to form a plug and so would need to undergo less preferential sliding. This is why, in Fig.~\ref{fig:Pf}(b), $\ell/\Lambda$ of plugged cases decreases and comes closer to that of open cases as the film thickness approaches the fastest-mode threshold. The increase in $P_f$ with film thickness [Fig.~\ref{fig:Pf}(a)] also results from humps on thicker films not having to undergo abnormally large sliding events in order to form plugs.

\begin{figure}
\centering
\includegraphics[width=0.8\textwidth]{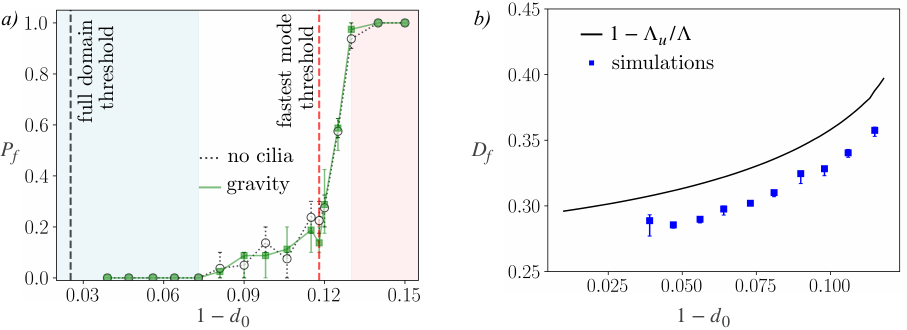}% Here is how to import EPS art
\caption{\label{fig:dryout_gravity}  (a) Plugging fraction for simulations run with (i) no ciliary transport ($u_c = 0$) and (ii) with weak gravitational acceleration instead of ciliary transport. (b) Fraction of the wall occupied by mucus-depleted zones: the measurements from the ensemble of simulations are compared with the fastest-mode based prediction (solid black line) of \citep{Swarnaditya-particles}.}
\end{figure}

\textit{Indifference to cilia and weak gravity.} In an axially symmetric problem, sliding arises as a perfect pitchfork bifurcation of the non-sliding evolution \citep{Dietze2018sliding} and is triggered by asymmetries in the random initial perturbation.
% and the ensuing sliding motion can occur in either direction with equal probability 
 The presence of axial asymmetry, owing to ciliary transport or weak gravitational forcing \citep{Dietze2015}, produces an imperfect bifurcation and hastens the onset of sliding \citep{Dietze2015}. 
% occurs more quickly and in the direction set by the asymmetry. So the humps always slide in the direction of ciliary transport.  imperfect The onset of the sliding instability is hastened by axial asymmetries. 
One may thus be tempted to think that ciliary transport causes intermediate films to plug with a higher probability. However, as discussed above, plugging is promoted by \textit{preferential} sliding of one hump, and so an effect that promotes the sliding of all humps equally will not increase the likelihood of plugging. We demonstrate this in Fig.~\ref{fig:dryout_gravity}(a), which shows that the plugging fraction is not appreciably decreased when ciliary transport is turned off [compare with Fig.~\ref{fig:Pf}(a)]. Reintroducing axial asymmetry via gravitational forcing also leaves the plugging statistics unchanged. As appropriate for small middle airways, we apply a weak forcing ($\mathcal{G} = 2 \times 10^{-4}$) that yields a mucus flow rate ten times slower than that produced by ciliary transport (for a flat film of thickness 0.117). Measurements of $\ell/\Lambda$ for these two scenarios are presented in the supplemental material \citep{supplement} and found to be analogous to those in Fig.~\ref{fig:Pf}(b). Moreover, we find that even the time required for plugging is insensitive to ciliary transport \citep{supplement}. So, sliding-induced plugging is not dependent on the presence of axial asymmetries such as ciliary transport or gravity. 

\textit{Mucus-depleted zones in open airways.} Before concluding, it is useful to measure the extent of the mucus depleted zones that are formed between unduloid-like humps, in open films. As shown in \citep{Swarnaditya-particles}, inhaled aerosols can deposit on these exposed sections of the airway wall---an outcome that is harmful in case of allergens and pathogens, and beneficial in case of aerosolized therapeutics. A simple theoretical estimate of the fraction $D_f$ of the wall covered by depleted zones is given by $(\Lambda-\Lambda_U)/\Lambda$, where $\Lambda_U$ is the width of the unduloid formed by the mucus volume contained in 
% a film of initial thickness $1-d_0$, spread uniformly across 
a unit cell of length $\Lambda$ (which is assumed to repeat identically across the whole airway of length $L$ owing to the dominance of the fastest-growing mode). 
 This prediction, calculated in \citep{Swarnaditya-particles}, is compared with measurements from our ensemble of long domain simulations in Fig.~\ref{fig:dryout_gravity}(b). [Ciliary transport is included here but we have found that it does not appreciably affect the results.] 
 % Because preferential sliding, and the consequent distortion of inter-hump spacing, is minimal in open films, 
We find rather good agreement. Importantly, the depleted zones are seen to widen with increasing mucus volume just as predicted in \citep{Swarnaditya-particles} (this counter-intuitive behavior is a consequence of the increasing dominance of radial over axial curvature as the film thickens).
% ---showing a counter-intuitive widening of depleted zones with increasing mucus volume.
% ---once again demonstrating the importance of the fastest-growing mode in setting the wavelength of the final array of humps. 
This agreement suggests that, when computing the deposition of aerosols on long airways, one can restrict attention to a periodic domain of length $\Lambda$ and still capture the distribution of mucus humps and depleted zones. 
\textit{ Conclusion.} So, when do mucus films plug airways? In short airways, which allow for just one growing hump, the transition between open and plugged airways is dictated by the limiting volume of equilibrium unduloids \citep{everett1972model}. 
 % The situation is more complex on long airways: 
 In long airways, however, the film gives rise to multiple humps whose ultimate fate is decided by the nonlinear dynamics of the film, with key roles played by the fastest-growing linear mode and the secondary sliding instability. 
 As a consequence, there is no simple transition from open to plugged airways, and the answer to when plugs form is probabilistic. \\
\textit{Acknowledgments.} We are grateful for financial support from
DST-SERB (J.R.P., grant no. SRG/2021/001185), the Indo–French Centre for the Promotion of Advanced Scientific Research (IFCPAR / CEFIPRA) (J.R.P. project no. 6704-A), and IRCC, IIT Bombay (S.H., Ph.D. fellowship; J.R.P., grant no. RD/0519-IRCCSH0-021).
J.R.P. also acknowledges his Associateship with the International Centre for Theoretical Sciences (ICTS), Tata Institute of Fundamental Research, Bangalore,
India. The authors thank the National PARAM Supercomputing Facility \textit{PARAM SIDDHI-AI} at CDAC, Pune for computing resources; simulations were also performed on the IIT Bombay workstations \textit{Gandalf} (procured through DST-SERB grant SRG/2021/001185), and \textit{Faramir} and \textit{Aragorn} (procured through the IIT-B grant RD/0519-IRCCSH0-021). 
%\end{acknowledgments}

\bibliography{mucus}% Produces the bibliography via BibTeX.

\end{document}